\documentclass[aps,preprint,floats,nofootinbib]{revtex4}
\usepackage{amsmath}
\usepackage{amssymb}
\usepackage{latexsym}
\usepackage{epsf}
\usepackage{epsfig}
\usepackage{graphicx}

\newlength{\defbaselineskip}
\newcommand{\setlinespacing}[1]%
           {\setlength{\baselineskip}{#1 \newcommand}}

\def\lsim{\mathrel{\raise.3ex\hbox{$<$\kern-.75em\lower1ex\hbox{$\sim$}}}} 
\def\gsim{\mathrel{\raise.3ex\hbox{$>$\kern-.75em\lower1ex\hbox{$\sim$}}}} 
\begin{document}

\preprint{
\hfill
\begin{minipage}[t]{3in}
\begin{flushright}
\vspace{0.0in}
FERMILAB--PUB--09--105--A\\
\end{flushright}
\end{minipage}
}

\hfill$\vcenter{\hbox{}}$

\vskip 0.5cm

\title{Astrophysical Uncertainties in the Cosmic Ray Electron and Positron Spectrum From Annihilating Dark Matter}
\author{Melanie Simet$^1$ and Dan Hooper$^{1,2}$}
\address{$^1$Department of Astronomy and Astrophysics, University of Chicago \\
$^2$Theoretical Astrophysics Group, Fermi National Accelerator Laboratory}
\date{\today}

\bigskip

\begin{abstract}

In recent years, a number of experiments have been conducted with the goal of studying cosmic rays at GeV to TeV energies.  This is a particularly interesting regime from the perspective of indirect dark matter detection. To draw reliable conclusions regarding dark matter from cosmic ray measurements, however, it is important to first understand the propagation of cosmic rays through the magnetic and radiation fields of the Milky Way. In this paper, we constrain the characteristics of the cosmic ray propagation model through comparison with observational inputs, including recent data from the CREAM experiment, and use these constraints to estimate the corresponding uncertainties in the spectrum of cosmic ray electrons and positrons from dark matter particles annihilating in the halo of the Milky Way.

\end{abstract}

\maketitle

\newpage

\section{Introduction}

In order for dark matter particles to be detected, they must either interact directly with particles of the standard model, or produce such particles through their annihilations.  In this paper, we concern ourselves with the indirect detection of dark matter through the observation of its annihilation products. If the dark matter consists of relic particles with weak-scale interactions and masses, they will produce a combination of  gamma rays, neutrinos, and other standard model particles in their annihilations.  Of particular observational interest are positrons~\cite{history} and antiprotons~\cite{antiprotons} which, in lieu of dark matter annihilations or other sources, are expected to be rare compared to their non-antimatter counterparts in the cosmic ray spectrum.

The most significant astrophysical source of cosmic ray positrons and antiprotons had generally been expected to be secondary production (creation through cosmic ray interactions occurring in the interstellar medium of the Milky Way). An excess of positrons or antiprotons in the cosmic ray spectrum relative to the small fraction expected as secondaries would imply the existence of primary sources of such particles. Although dark matter annihilations may be capable of producing such a signature~\cite{history,BaltzEdsjo,HooperSilk}, astrophysical sources have also been proposed, including pulsars~\cite{pulsar1}. Very recently, it has also been suggested that secondary production of positrons {\it within} the regions of cosmic ray acceleration may also produce a significant fraction of the cosmic ray positron flux at high energies~\cite{Blasi:2009hv}. In order to distinguish between such sources of cosmic ray antimatter, a detailed understanding of the processes involved in cosmic ray propagation is likely to be required.

The satellite-based experiment PAMELA (a Payload for Antimatter Matter Exploration and Light-nuclei Astrophysics) has reported a steadily increasing positron fraction (the ratio of positrons to positrons-plus-electrons) in the cosmic ray spectrum between approximately 10 GeV and 100 GeV~\cite{PAMELA} (data has not yet been published at higher energies). While consistent with previous indications from the HEAT~\cite{heat} and AMS-01~\cite{ams01} experiments, this result is in stark contrast to the behavior expected if the positron spectrum were dominated by secondary particles produced during cosmic ray propagation~\cite{secondaries}. Although the origin of these particles is not yet known, their observation suggests the existence of a primary source (or sources) of positrons~\cite{serpico}. PAMELA's measurement of the antiproton-to-proton ratio, in contrast, is consistent with purely secondary production~\cite{pamelaantiproton}. The antiproton result from PAMELA can be used to limit the nature of positron primary sources, including on the range of dark matter models potentially responsible~\cite{antiprotonconst}. PAMELA is ultimately expected to measure the spectra of cosmic ray protons, antiprotons, electrons, and positrons up to energies of 700 GeV, 190 GeV, 2 TeV, and 270 GeV, respectively.

ATIC (Advanced Thin Ionization Calorimeter) is a balloon-based experiment designed to study the spectra of cosmic ray protons, light nuclei, and electrons. Recently, the ATIC collaboration published their electron (plus positron, as they do not distinguish between these species) spectrum between approximately 20 GeV and 2 TeV~\cite{ATIC}. Remarkably, they find that the spectrum contains a bump-like feature over the steadily declining power-law between roughly 300 and 800 GeV, peaking at around 600 GeV. The energy loss and diffusion rates of electrons in this energy range lead us to conclude that these particles must originate within approximately $\sim$1 kpc of the Solar System.

In light of these recent excesses observed in the cosmic ray positron fraction and electron (plus positron) spectrum, we revisit this topic, concentrating on the astrophysics that goes into determining the cosmic ray electron and positron spectra resulting from annihilating dark matter. After an initial spectrum of electrons and positrons is created through dark matter annihilations, that spectrum evolves as it diffuses and propagates through the radiation fields and magnetic field of the Milky Way. The properties of the propagation model used to describe this evolution can be constrained from other cosmic ray observations, including the relative abundances of unstable and stable secondaries in the cosmic ray spectrum, which provide us with information pertaining to the characteristic timescales over which the particles have been propagating and the integrated density of matter through which it has passed. Together, such measurements enable us to construct a reasonably constrained parameterization of the propagation model, which can then be used to calculate, among other things, the propagated spectrum of cosmic ray electrons and positrons from dark matter annihilations.

The remainder of this paper is structured as follows. In Section~\ref{propagation}, we use a large array of cosmic ray nuclei data in conjunction with simulations conducted using the GALPROP program (v50p)~\cite{Strong} to constrain the characteristics of the model describing cosmic ray propagation in our galaxy.  In Section~\ref{dm}, we apply this constrained model to the electrons and positrons produced in dark matter annihilations and compare our results to the positron fraction measured by PAMELA. We summarize our results in Section~\ref{conclusion}. 

\section{Constraining the Propagation Model}
\label{propagation}

To model the diffusion, nuclear interactions, and energy loss processes of galactic cosmic rays, we use the publicly available GALPROP code~\cite{Strong}.  This code begins by injecting cosmic rays with a supernova-like isotope distribution and a parameterized spatial distribution chosen to reproduce the EGRET data.  The individual isotopes, starting with the largest atomic number, are then propagated through the galaxy, with the energy distribution and composition altered through a combination of spacial diffusion, energy losses, diffusion in momentum space (diffusive reacceleration), electron K-capture, convection, spallation, and radioactive decay.

These processes are collectively described by the cosmic ray propagation equation. For particles of momentum $p$ with particle density per unit momentum $\psi(\vec{x},p,t)$, this equation is given by~\cite{Strong}:
\begin{eqnarray}
\frac{\partial\psi(\vec{x},p,t)}{\partial t} = q(\vec{x},p) &+&\vec{\nabla}\cdot [D_{xx}\vec{\nabla}\psi(\vec{x},p,t) -\vec{V_c}\psi(\vec{x},p,t)]+\frac{\partial}{\partial p}p^2 D_{pp} \frac{\partial}{\partial p}\frac{1}{p^2}\psi(\vec{x},p,t) \nonumber \\
 &-& \frac{\partial}{\partial p}[\dot{p}\psi(\vec{x},p,t) - \frac{p}{3}(\vec{\nabla}\cdot\vec{V_c})\psi(\vec{x},p,t)]-\frac{1}{\tau_f}\psi(\vec{x},p,t)-\frac{1}{\tau_r}\psi(\vec{x},p,t) 
\label{diffusionloss} 
\end{eqnarray}
where $D_{xx}$ is the diffusion constant, which we parameterize by
\begin{equation}
D_{xx}=\beta \, D_{0xx}\left(\frac{\rho}{4 \mbox {GV}}\right)^{\alpha},
\end{equation} 
where $\beta$ is the particle's speed and $\rho$ is its rigidity (momentum per unit charge). The diffusion constant describing reacceleration, $D_{pp}$, is related to $D_{xx}$ by~\cite{pp,berebook}
\begin{equation} 
D_{pp}=\frac{4p^2v_A^2}{3\alpha(4-\alpha^2)(4-\alpha)} \frac{1}{D_{xx}},
\end{equation}
where $v_A$ the Alfv\'{e}n speed. In the propagation equation (Eq.~\ref{diffusionloss}), $V_c$ is the convection velocity, $\tau_f$ is the fragmentation time, $\tau_r$ is the radioactive decay time, and $q(\vec{x},p)$ is the source term.  The source term includes not only the injection spectrum, but also the products of the decay and spallation of heavier species of nuclei.  The equation is solved assuming a cylindrical geometry over the diffusive region (a volume with a half-thickness of $L_{eff}$ and a radius of 20 kpc).  Outside of this volume, the particles are not confined by the Galactic Magnetic Field and freely escape. Assuming that the system is near steady state equilibrium, we set the left side of Eq.~\ref{diffusionloss} to zero in solving for $\psi(\vec{x},p,t)$.

Note that we refer to the half-thickness of the diffusive region as $L_{eff}$ rather than simply as $L$. Although GALPROP treats the diffusion constant to be the same in all directions (and locations throughout the diffusion zone), the actual process of diffusion in the galaxy is the result of magnetic fields which are not spherically symmetric, but are instead structured and thought to be disk-like in form. The quantity $L_{eff}$ thus does not denote simply the physical thickness of the diffusion zone, but may also include information regarding the differing efficiencies of vertical and horizontal diffusion.

In the course of this study, we completed 370 simulations with GALPROP. In these runs, we varied four parameters: the normalization of the diffusion coefficient ($D_{0xx}$), the slope of the dependence of the diffusion coefficient with rigidity ($\alpha$), the effective half-thickness of the diffusion region ($L_{eff}$), and the convection velocity ($V_c$).  We considered values of $D_{0xx}$ between $1.08\times 10^{28}$ and $2.42\times 10^{29}$ $\mbox{cm}^2 \, \mbox{s}^{-1}$ (at a reference rigidity of 4 GeV),  $\alpha$ between 0.34 to 0.52 and $L_{eff}$ between 1 and 17 kpc.  $V_c$ was varied between 0 and 15 km/s/kpc, where we have assumed that the convection velocity scales proportionally to the distance from the Galactic Plane.  We have adopted a two-dimensional (cylindrical) symmetry which allows physical quantities to vary with $R$ and $z$, but not with $\theta$.  Unless stated, all other parameters are unchanged from those found in the GALPROP definitions file (\texttt{galdef\_50p\_599278})~\cite{galpropweb}. In particular, note that we have not changed the Alfv\'{e}n speed $V_{a}$ from its default value of 36 km/s.  Some previous studies (for example, see Ref.~\cite{alf}) have allowed $V_a$ to vary and have found a fairly large range of values to be consistent with cosmic ray data, so long as compensating changes are made in other propagation parameters accordingly.  With the recent introduction of new high energy data from the CREAM experiment~\cite{CREAM}, however, this parameter is more tightly constrained.

Once the GALPROP simulations were completed, the spectra of the various elements and isotopes at the location of the Solar System ($R=8.5$ kpc, $z=0$) were extracted and combined to yield predictions for both stable and unstable secondary-to-primary ratios. Stable secondary-to-primary ratios, such as B/C and sub-Fe/Fe, are valuable measures of the average amount of matter traversed by cosmic rays as a function of energy. Although B/C is especially useful, sub-Fe/Fe can provide complementary information, as it is most sensitive to a slightly different range of energies. Unstable secondary-to-primary ratios, in contrast, serve as a measurement of the time cosmic rays have been propagating. Beryllium-10 is particularly useful in this regard, being the longest lived and best measured unstable secondary. The measurement of $^{10}$Be/$^{9}$Be serves as a clock, since the ratio of the radioactive isotope to the stable one is directly related to the amount of time elapsed since the creation of the particles. For an excellent review of this and related topics, see Ref.~\cite{StrongMoskPtusk}.



\begin{table}
\begin{tabular}{l | c | c}
 Authors & Ratios & Number \\
 \hline
  \citet{CREAM} (CREAM) & B/C & (1) \\
  \citet{Panov} (ATIC) & B/C & (2) \\
  \citet{de Nolfo}$^a$ (ISOMAX)& $^{10}$Be/$^9$Be & (3)\\
  \citet{Engelmann} (HEAO-3)  & B/C, sub-Fe/Fe & (4) \\
  \citet{Swordy} (Spacelab 2) & B/C & (5)\\
  \citet{Dwyer}$^a$ & B/C, sub-Fe/Fe & (6)\\
  \citet{Webber85}$^a$ & B/C & (7) \\  
  \citet{Chappell} & B/C & (8)\\
  \citet{Simon} & B/C & (9)\\
  \citet{Orth} & B/C & (10)\\
  \citet{Caldwell} (IMP-8)& B/C & (11) \\
  \citet{Juliusson} & B/C & (12)
\end{tabular}
\caption{The cosmic ray data used to constrain the propagation model. The superscript ``{\it a}'' denotes papers whose authors state that there may be additional systematic errors, but do not characterize them.}
\label{RatioData}
\end{table}

%
\begin{table}
\begin{tabular}{l | c | c}
 Authors & Ratios & Number \\
 \hline
  \citet{Webber03}$^b$ (Voyager) & B/C, sub-Fe/Fe & (13) \\
  \citet{Webber02}$^b$ (Voyager) & $^{10}$Be/$^9$Be & (14)\\
  \citet{Hams}$^b$ (Ulysses) & $^{10}$Be/$^9$Be & (15) \\
  \citet{Davis}$^b$ (CRIS) & B/C, sub-Fe/Fe & (16) \\
  \citet{Yanasak}$^b$ (CRIS) & $^{10}$Be/$^9$Be & (17) \\
  \citet{Connell}$^b$ (Ulysses) & $^{10}$Be/$^9$Be &  (18) \\
  \citet{DuVernois}$^b$ (Ulysses) & B/C, sub-Fe/Fe & (19) \\
  \citet{Leske}$^b$ (ISEE-3) & sub-Fe/Fe & (20) \\
  \citet{Esposito}$^b$ (ALICE) & sub-Fe/Fe & (21) \\
  \citet{Krombel}$^b$ (ISEE 3) & B/C & (22)\\
  \citet{G-M87}$^b$ (IMP-8) & B/C & (23)\\
  \citet{G-M81}$^b$ & $^{10}$Be/$^9$Be & (24) \\
  \citet{Young}$^b$ & sub-Fe/Fe & (25)\\
  \citet{Wiedenbeck}$^b$ & $^{10}$Be/$^9$Be & (26) \\
  \citet{Webber79}$^b$ & sub-Fe/Fe  & (27) \\
  \citet{Buffington}$^b$ & $^{10}$Be/$^9$Be & (28) \\
  \citet{Lezniak}$^b$ & B/C & (29) \\
  \citet{Hagen}$^b$ & B/C, 10Be/9Be & (30)\\
  \citet{Maehl}$^b$ & B/C & (31)\\
  \citet{Benegas}$^b$ & sub-Fe/Fe & (32) \\
  \citet{Lund}$^b$ & B/C & (33)\\
%
\end{tabular}
\caption{Cosmic ray data at energies below the range we have considered in our analysis.}
\label{RatioDatalow}
\end{table}

For each set of diffusion parameters, we have compared the predicted secondary-to-primary ratios to the body of current observational data, as listed in Table~\ref{RatioData}. We have limited the data included in our analysis to energies above 5 GeV for the B/C and sub-Fe/Fe ratios, and above 1 GeV for the $^{10}$Be/$^9$Be data (no data above 5 GeV is currently available for this ratio). At lower energies, the effects of solar modulation become increasingly important, making it difficult to reliably compare the data to the predictions of a given propagation model. We also list in Table~\ref{RatioDatalow} a collection of cosmic ray data taken only at energies below those we have considered in our analysis. In the case of the $^{10}$Be/$^9$Be measurement at 1-2 GeV, in an effort to lessen the impact of solar modulation, we have included the correction as described in Ref.~\cite{de Nolfo}.

\begin{figure}
\begin{center}
\begin{tabular}{c}
\hspace{-1.3cm}
\epsfig{file=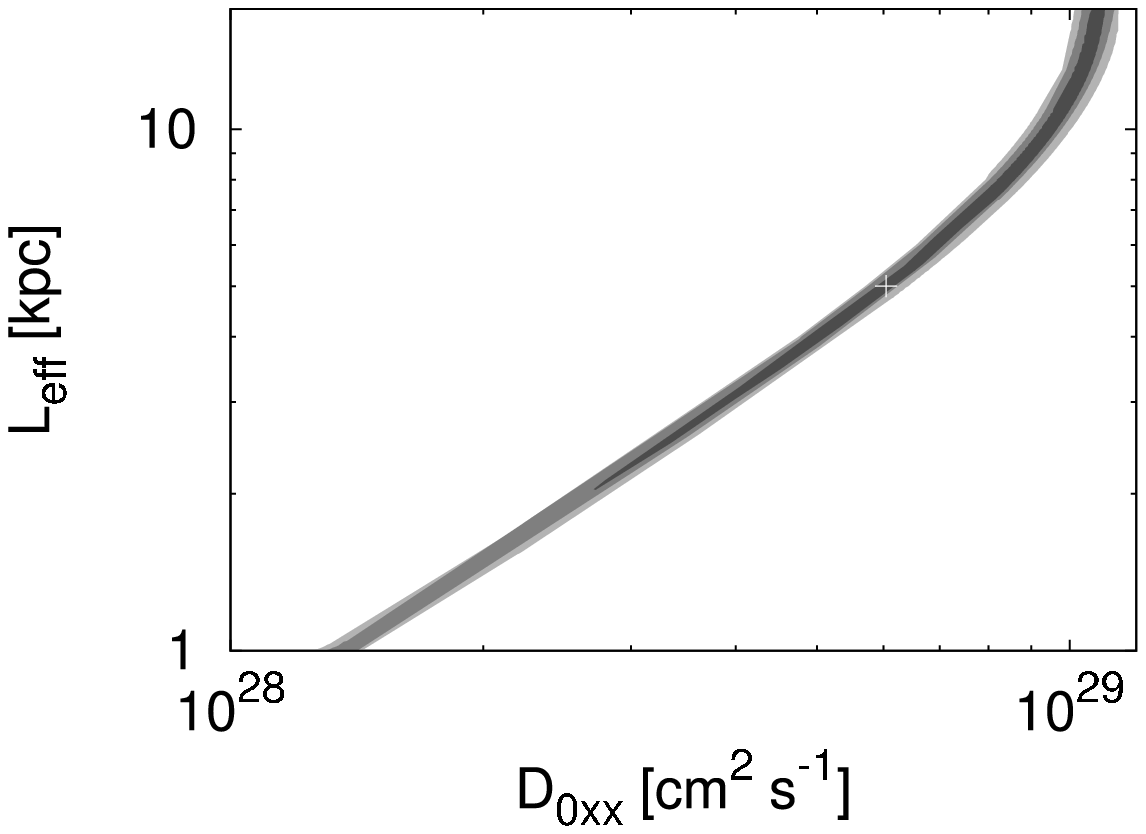,width=0.55\columnwidth}
\hspace{-1.0cm}
\epsfig{file=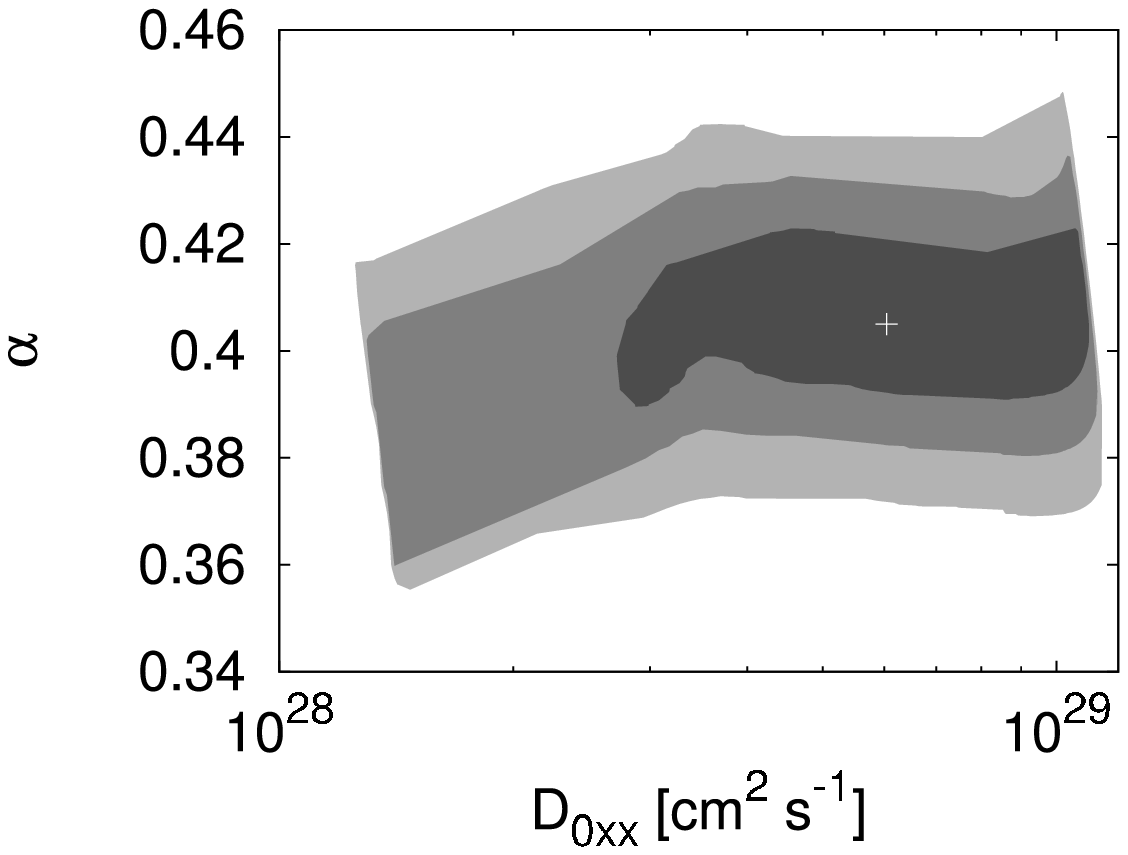,width=0.55\columnwidth}
\vspace{-0.5cm}\\
\hspace{-2.0cm}
\epsfig{file=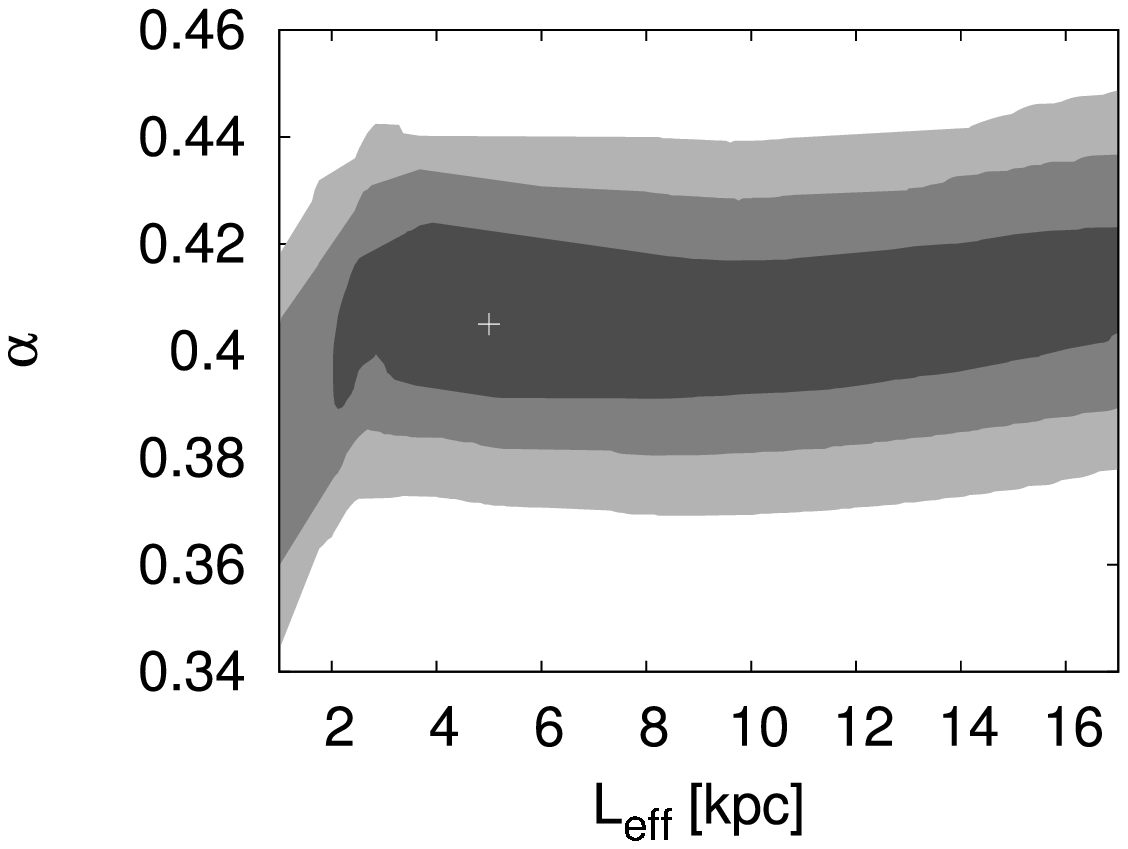,width=0.55\columnwidth}
\end{tabular}
\end{center}
\vspace{-1.5cm}
\caption{The (1, 2, and 3$\sigma$) regions of the propagation parameters, calculated with respect to the data listed in Table~\ref{RatioData}.  Results are marginalized over $\alpha$, $L_{eff}$, and $D_{0xx}$, respectively.  The white crosses mark the best fit model in each frame. In this figure, the effects of convection have been neglected.}
\label{v0}
\end{figure}

In Fig.~\ref{v0}, we plot the acceptable ranges of the propagation model parameters, based on our $\chi^2$ calculation. Neglecting the effects of convection for the time being, we find the overall best fit for the following set of propagation parameters: $D_{0xx}=6.04\times 10^{28}$ cm$^{2}$ s$^{-1}$, $L_{eff}=5.0$ kpc, and ${\alpha}=0.41$, which yields a $\chi^2$ per degree-of-freedom of 1.37. Although the acceptable parameter regions are well constrained in $D_{0xx}$ and $\alpha$, the allowed values for $L_{eff}$ extend beyond the range we considered (1-17 kpc). For physical reasons, however, we do not consider values outside of this range. We find that the data prefer values of $\alpha$ which lie between those predicted for Kolmogorov-type ($\alpha=1/3$)~\cite{kolmogorov} and Kraichnan-type ($\alpha=1/2$)~\cite{kraichnan} turbulence.



\begin{table}
\begin{center}
\begin{tabular}{l |  c || c | c | c}
 \hline
  &Model & A1 & A2 & A3\\
 \hline
  &$D_{0xx}$ (cm$^2$/s) & $2.72 \times 10^{28}$ & $5.47 \times 10^{28}$ & $1.10 \times 10^{29}$ \\
 1$\sigma$ &  $L_{eff}$ (kpc) & 2.04 & 4.52 & 17.0 \\
  & $\alpha$ & 0.39 & 0.42 & 0.41 \\
  \hline
   & Model & A4 & A5 & A6 \\
  \hline
  & $D_{0xx}$ (cm$^2$/s) & $1.29\times 10^{28}$ & $4.02\times 10^{28}$ & $1.13\times 10^{29}$ \\
  2$\sigma$ & $L_{eff}$ (kpc) & 1.0 & 3.0 & 17.0 \\
   & $\alpha$ & 0.40 & 0.44 & 0.39 \\
 \hline
\end{tabular}
\end{center}
\caption{Selected (extrema and central value) propagation model parameters which have been found to be consistent with the cosmic ray nuclei data at the 1$\sigma$ (A1, A2 and A3) and 2$\sigma$ (A4, A5 and A6) levels, neglecting the effects of convection.}
\label{Models}
\end{table}

\begin{figure}
\begin{center}
\hspace{-1.0cm}
\epsfig{file=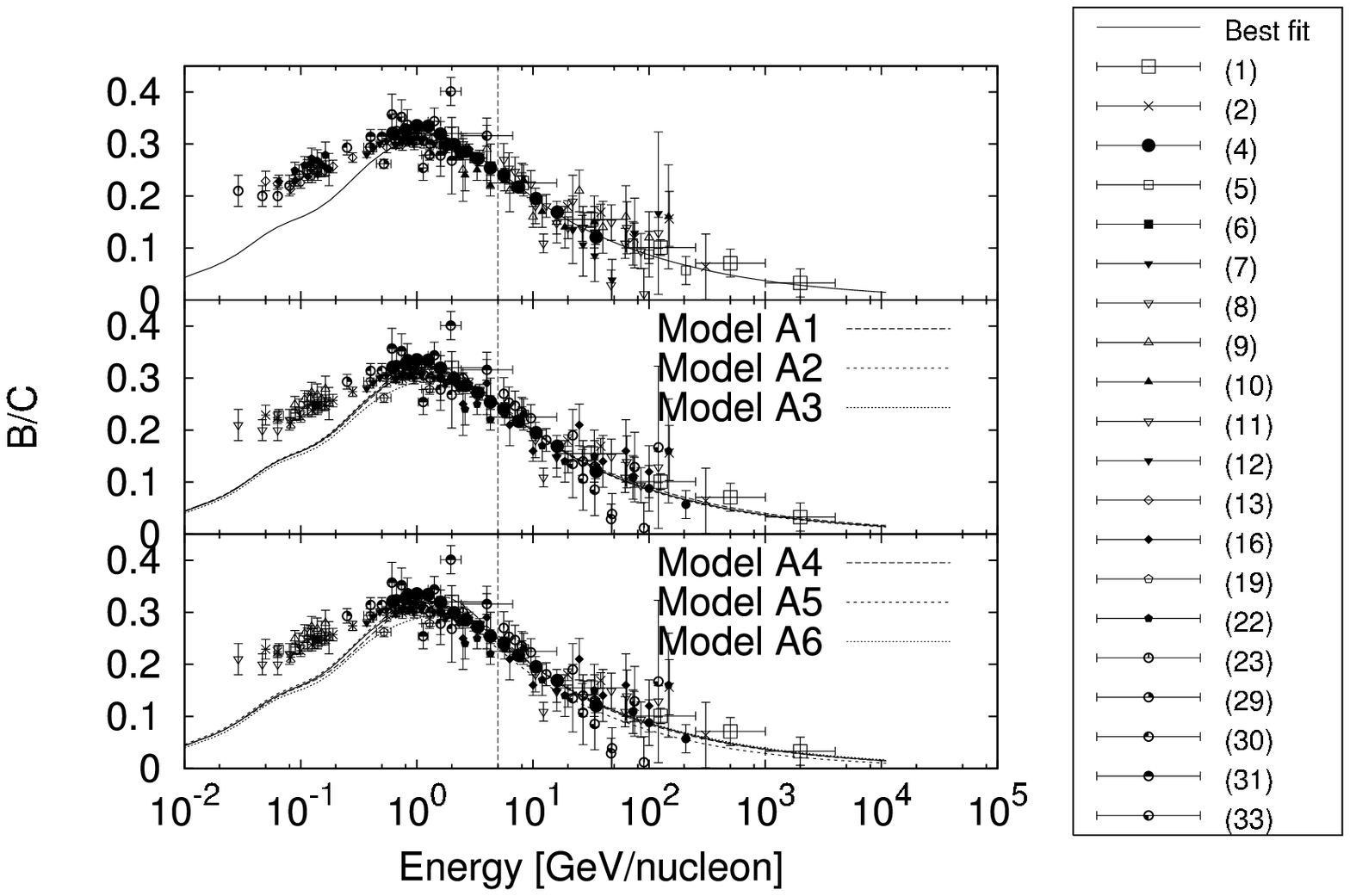,width=1.0\columnwidth} \\
\hspace{-1.0cm}
\epsfig{file=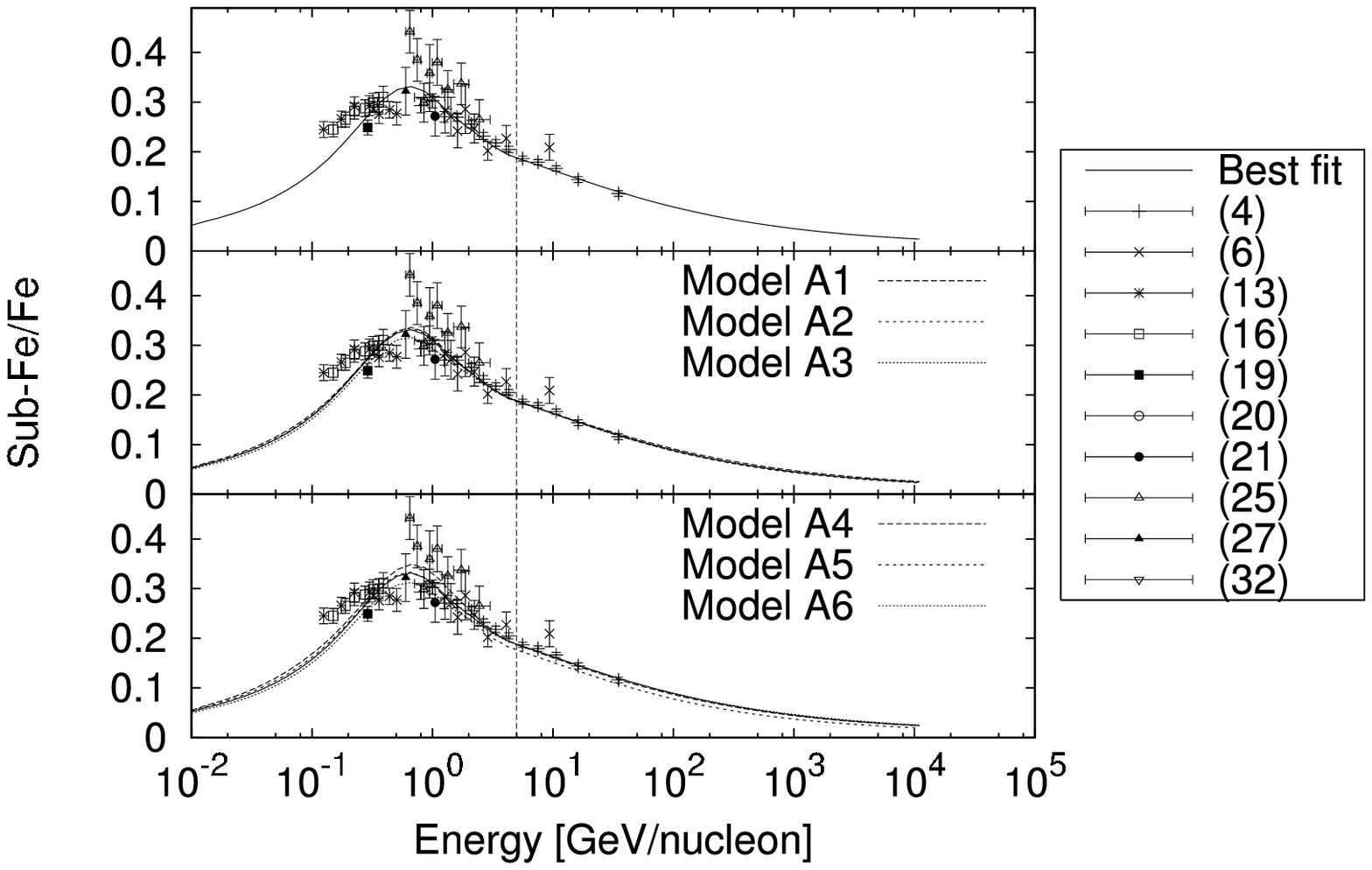,width=1.0\columnwidth} 
\end{center}
\vspace{-1.0cm}
\caption{Predictions of the propagation model described in the text compared to current measurements of the boron-to-carbon ratio (top) and the sub-iron (Sc+Ti+V)-to-iron ratio (bottom) in the cosmic ray spectrum. The model parameters used are those described in Table~\ref{Models} and in the text. The data shown is listed in Tables~\ref{RatioData} and \ref{RatioDatalow}, as indicated by the key.  In calculating the quality of the fit of a propagation model, only data to the right of the vertical line (greater than 5 GeV) has been included.}
\label{RatioSpectraBC}
\end{figure}

\begin{figure}
\begin{center}
\hspace{-1.0cm}
\epsfig{file=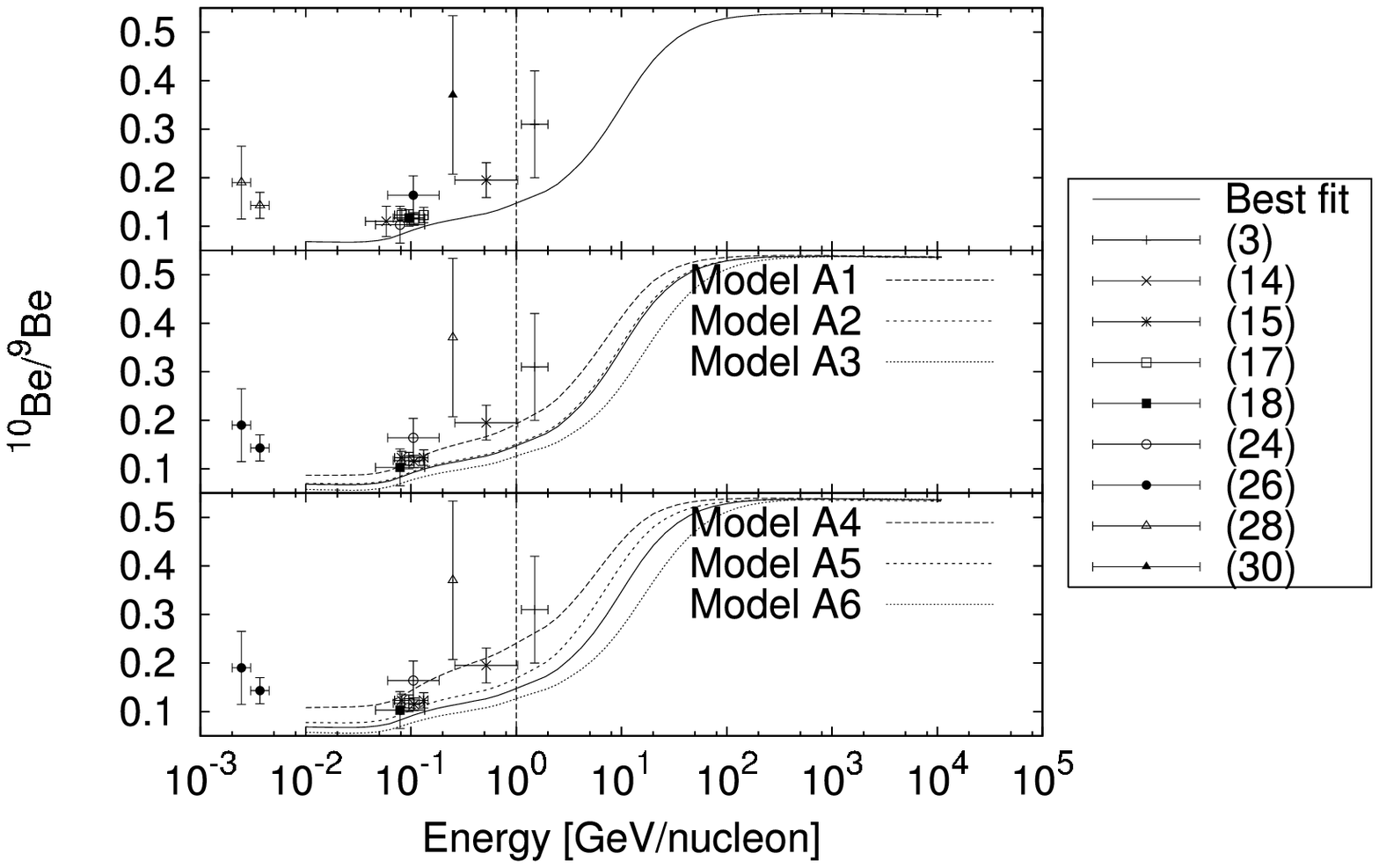,width=1.0\columnwidth} \\
\end{center}
\vspace{-1.0cm}
\caption{Predictions of the propagation model described in the text compared to current measurements of the beryllium 10-to-beryllium 9 ratio in the cosmic ray spectrum. The model parameters used are those described in Table~\ref{Models} and in the text. The data shown is listed in Tables~\ref{RatioData} and \ref{RatioDatalow}, as indicated by the key.  In calculating the quality of the fit of a propagation model, only data to the right of the vertical line (the single highest energy error bar) has been included.}
\label{RatioSpectraBe}
\end{figure}

In Table~\ref{Models}, we list the parameters for a selection of extrema and central value propagation models.  The B/C, sub-Fe/Fe and $^{10}$Be/$^9$Be ratios predicted in these models, along with those predicted in the best fit model, are compared to the cosmic ray nuclei data in Figs.~\ref{RatioSpectraBC} and~\ref{RatioSpectraBe}. These figures confirm that this range of propagation models provides a reasonably good fit to the current set of cosmic ray nuclei data over the energy range considered.

To illustrate the effect of changing the Alfv\'{e}n speed, we performed additional GALPROP runs with $V_a=72$ km/s, $D_{0xx}=4.31\times 10^{28}\,\mbox{cm}^{2} \, \mbox{s}^{-1}$, and a range of values for $L_{eff}$ and $\alpha$.  The best $\chi^2$ per degree-of-freedom found was 1.96, which is considerably larger than the 1.39 found with our default value ($V_a=36$ km/s). From this test, we conclude that large variations in the Alfv\'{e}n speed relative to our default choice are disfavored by the current data set.

\begin{figure}
\begin{center}
\hspace{-2.0cm}
\epsfig{file=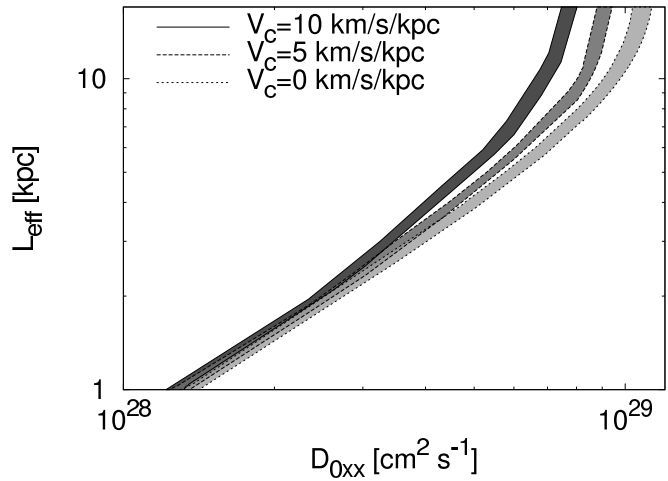,width=0.6\columnwidth}\\
\end{center}
\vspace{-1.7cm}
\caption{The 2$\sigma$ region in the $D_{0xx}$, $L_{eff}$ plane, with and without the effect of convection, $V_c$ = 0, 5, and 10 km/s/kpc.}
\label{v05}
\end{figure}

Thus far, our analysis has neglected the effects of convection (preferential motion of cosmic rays away from the Galactic Plane). To explore how convection effects our results, we performed several further runs of GALPROP. From these runs, we learned that the fractional change in the element ratios due to convection was approximately of the same shape across all values of the convection velocity, $V_c$, at least up to 15 km/s/kpc, with an overall normalization varying linearly with the velocity.  The ratios with convection were thus estimated by calculating the fractional change with $V_c$=15 km/s/kpc, and then multiplying the interpolated result with the matching fractional change scaled by $V_c/(15$ km/s/kpc).  2$\sigma$ regions in the $L_{eff}-D_{0xx}$ plane are given for $V_c$=0, 5, and 10 km/s/kpc in Fig.~\ref{v05}. This illustrates how convection could potentially alter the allowed range of the propagation parameters.


Before turning our attention to the propagation of electrons and positrons from dark matter annihilations, we would like to make some general remarks regarding the methods used in this section. Beginning with the cosmic ray propagation model as described by Eq.~\ref{diffusionloss}, we have found a range of parameters which yield good agreement with the current cosmic ray data at energies high enough to be only modestly impacted by solar modulation. Although this propagation model contains a wide variety of physical effects, including spatial diffusion, energy losses, diffusive reacceleration, electron K-capture, convection, spallation, and radioactive decay, it does have its limitations. For example, we have implicitly assumed that the diffusion constant, $D_{xx}$, does not vary with location (within the boundary conditions). Although such simplifying approximations are currently necessary to make the problem of constraining the propagation model tractable, we can be certain that at some point in the future this model will break down (fail to adequately describe the observations) and require a more sophisticated treatment. As our best fit parameter sets provide reasonable fits to the current data, however, it is appears that the approach described here is a reasonably accurate, or at an least adequate, approximation of the behavior of cosmic ray propagation in the Milky Way.


\clearpage

\section{High Energy Electrons And Positrons From Dark Matter Annihilations In the Milky Way}\label{dm}

In this section, we apply the constraints on the cosmic ray propagation model obtained in the previous section to the problem of cosmic ray electron and positron propagation. We begin by adapting Eq.~\ref{diffusionloss} to the case of high energy electrons and positrons. In particular, we remove the terms describing fragmentation and radioactive decay. Furthermore, we focus our analysis uniquely on cosmic ray electrons and positrons with energies above $\sim$\,10 GeV, at which the effects of reacceleration and convection are expected to be negligible. In the steady state limit, the propagation equation for high energy electrons and positrons reduces to 
\begin{eqnarray}
0 = q(\vec{x},p) +\vec{\nabla}\cdot [D_{xx}\vec{\nabla}\psi(\vec{x},p,t)]
 + \frac{\partial}{\partial p}[B(p)\, \psi(\vec{x},p,t)],  
\end{eqnarray}
where $B(p)$ is the energy loss rate of electrons due to synchrotron and inverse Compton processes. In the relativistic limit, this rate is related to the radiation field and magnetic field energy densities by
\begin{eqnarray}
B(E_e) &=& \frac{4}{3}\sigma_T \rho_{\rm rad} \bigg(\frac{E_e}{m_e}\bigg)^2    + \frac{4}{3} \sigma_T \rho_{\rm mag} \bigg(\frac{E_e}{m_e}\bigg)^2 \nonumber \\
&\approx& 1.02 \times 10^{-16}\,{\rm GeV/s}\,\bigg(\frac{\rho_{\rm rad}+\rho_{\rm mag}}{{\rm eV}/{\rm cm}^3}\bigg) \times  \bigg(\frac{E_e}{{\rm GeV}}\bigg)^2 \nonumber \\
&\equiv& \frac{1}{\tau} \times \frac{E^2_e}{(1\,{\rm GeV})},
\end{eqnarray}
where $\sigma_T$ is the Thompson cross section and $\tau$ is the representative energy loss time. There is also a contribution to the energy loss rate due to Bremsstrahlung, but this is significant only at energies lower than those considered here.

\begin{figure}
\begin{center}
\hspace{-2.0cm}
\epsfig{file=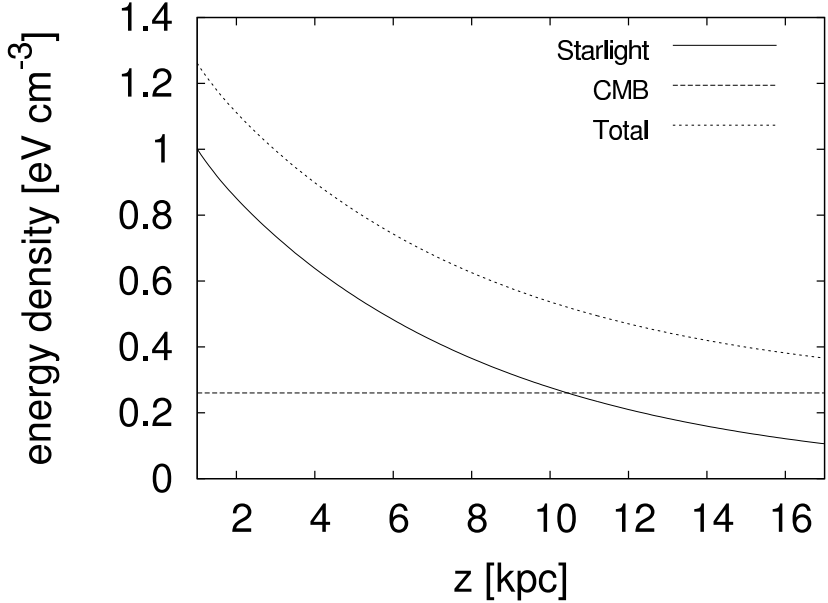,width=0.75\columnwidth}
\end{center}
\vspace{-2.0cm}
\caption{The energy density of radiation at $R=8.5$ kpc as a function of the distance away from the Galactic Plane, $z$.  The starlight density out to $z=5$ kpc is as is implemented in the GALPROP program. The density beyond $z=5$ kpc was estimated by extrapolation. The energy density of the Galactic Magnetic Field is not included. See text for more details.}
\label{uphoton}
\end{figure}

The radiation fields contributing to the inverse Compton loss rate include starlight, emission from dust, and the cosmic microwave background. In Fig.~\ref{uphoton}, we show the density of the interstellar radiation field (ISRF) as estimated in Ref.~\cite{ISRF} at a distance of $R=8.5$ kpc from the Galactic Center as a function of the distance away from the Galactic Plane, along with the uniform cosmic microwave background density. In addition, the magnetic field energy density which leads to synchrotron losses is related to the RMS field strength by $\rho_{\rm mag}=B^2/2\mu_0$. For $B\approx 3\, \mu$G (5 $\mu$G) in the local Milky Way, we arrive at a reasonable estimate of $\rho_{\rm mag} \approx 0.2$ eV/cm$^3$ (0.6 eV/cm$^3$). 

As we did for the diffusion constant, we perform our analysis under the approximation that the radiation and magnetic field densities do not have a strong spatial dependence within the relevant volume of the diffusion zone. Considering the distribution shown in Fig.~\ref{uphoton}, the radiation field density varies by only $\sim$10\% ($\sim$30\%) out to distances of 2 kpc (4 kpc) away from the Galactic Plane. As high energy electrons and positrons originating from greater distances are expected to lose most of their energy before reaching the Solar System, the spatial variation of the energy loss rate is not expected to significantly impact the observed spectrum, except perhaps at the lowest energies considered here ($\sim$\,10-20 GeV).

\begin{figure}
\begin{center}
\begin{tabular}{c}
\epsfig{file=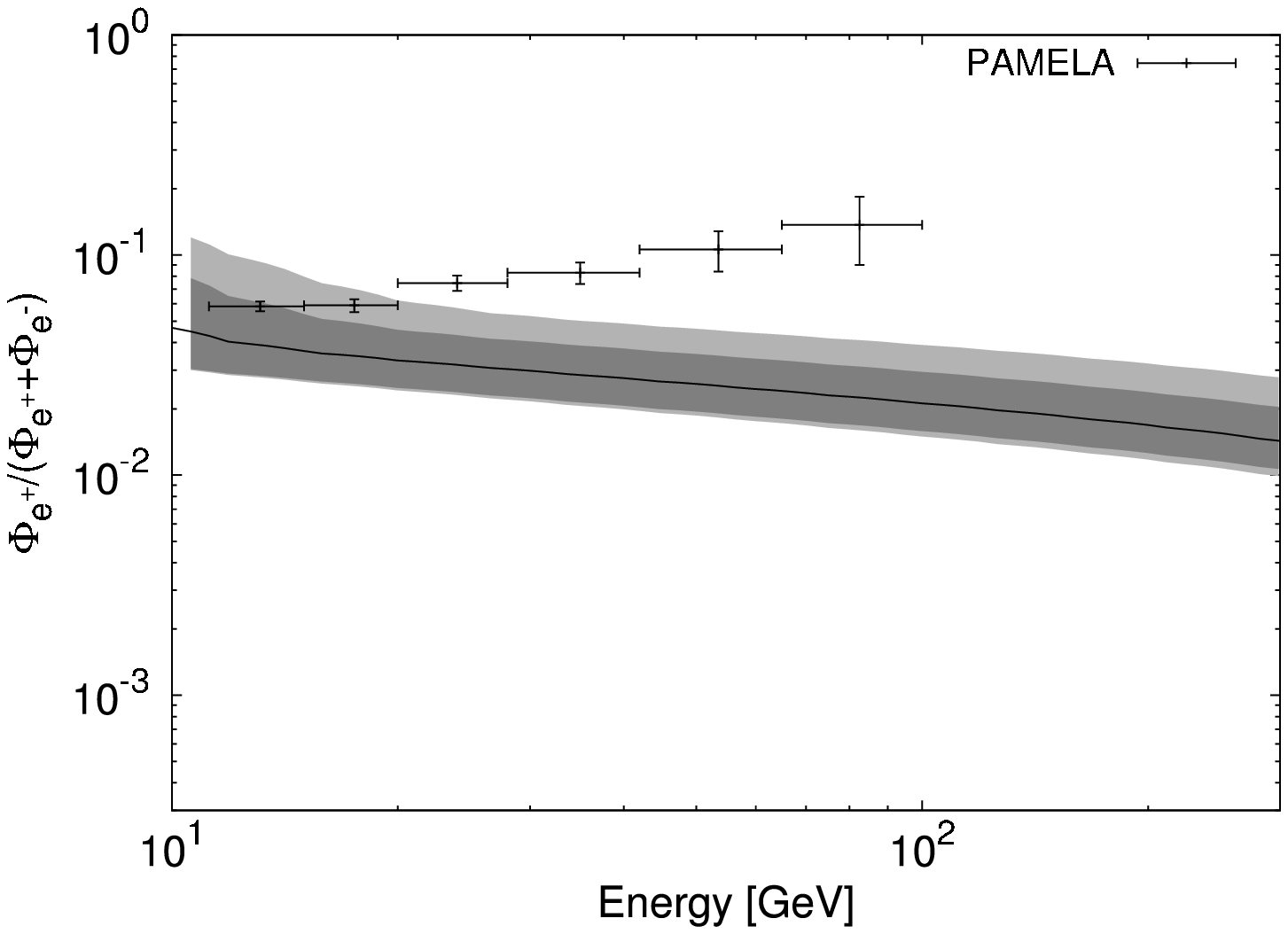,width=0.7\columnwidth} \\
\epsfig{file=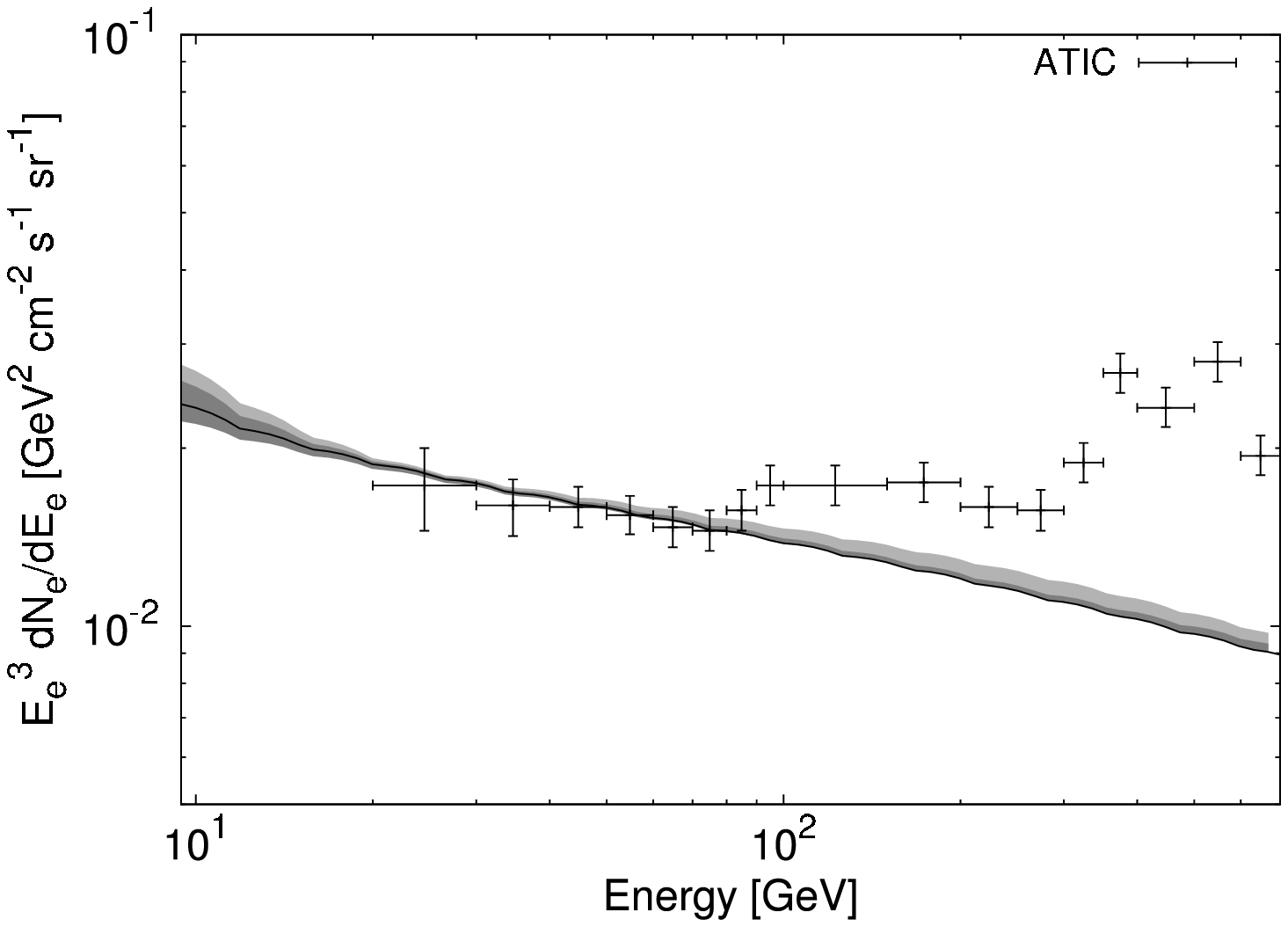,width=0.7\columnwidth} \\
\end{tabular}
\end{center}
\caption{The positron fraction (top) and electron-plus-positron spectrum (bottom) predicted for the range of propagation parameters which provide acceptable fits to the cosmic ray data, assuming an injected electron spectrum described by $dN_e/dE_e \propto E_e^{-2.5}$. The solid line denotes the result using the best fit propagation model, whereas the dark gray and light gray regions describe the results found over the 1 and 2$\sigma$ range of propagation models (see Fig.~\ref{v0}). Shown for comparison are data from the PAMELA~\citep{PAMELA} and ATIC~\citep{ATIC} experiments. It is clear that, without the inclusion of an additional source of cosmic ray positrons/electrons, these models are inconsistent with the results of these experiments.}
\label{comparebands}
\end{figure}

In this section, we consider the range of propagation parameters shown in Fig.~\ref{v0} (the 1$\sigma$ and 2$\sigma$ contour regions), which collectively represent the range of propagation models found to be consistent with the cosmic ray data. Even without considering any primary sources of cosmic ray positrons, the variation over this range of propagation parameters leads to a range of predicted cosmic ray electron (primary plus secondary) and positron (secondary) spectra.  Beginning with an injected (primary) spectrum of electrons described by $dN_e/dE_e \propto E_e^{-2.5}$, we have used GALPROP to calculate the resulting electron and positron spectra after propagation for each of the acceptable parameter sets. For this injected electron spectrum, each of the parameter sets leads to an electron spectrum at the Solar System with a slope over the range of 5-100 GeV of $dN_e/dE_e \propto E^{-\delta}$, where $\delta \approx 3.2-3.3$, which is in reasonable agreement with the measured cosmic ray electron slope~\cite{electronspec}. The predicted positron fraction over this energy range, as well as the electron spectrum at higher energies, are not in agreement with the observations of the PAMELA or ATIC experiments, however. The predicted positron fraction (top) and electron-plus-positron spectrum (bottom) are shown in Fig.~\ref{comparebands}, compared to the data from PAMELA and ATIC. It is clear that none of these propagation models lead to an acceptable fit to the data. Another source of cosmic ray positrons/electrons is required to accommodate these observations.

Turning our attention now to the source term in the propagation equation, we assume that dark matter is distributed throughout the galaxy following an Navarro-Frenk-White (NFW) halo density profile~\cite{NFW},
\begin{equation}
\rho(r,z) = \rho_0\frac{(8.5/20)[1+(8.5/20)]^2}{(\sqrt{r^2+z^2}/20\mbox{ kpc})[1+(\sqrt{r^2+z^2}/20 \mbox{ kpc})]^2},
\end{equation}
with a local dark matter density, $\rho_0$, of 0.3 GeV/cm$^{3}$.  The uncertainty introduced by limiting ourselves to such a profile is modest for the energy range we are considering (see Ref.~\cite{HooperSilk} for a comparison of different halo profiles). For example, the electron/positron spectrum resulting from a highly cusped Moore {\it et al.} profile~\cite{moore} or an isothermal sphere profile are virtually indistinguishable from that predicted for an NFW profile above $\sim$40 GeV, and vary by less than a factor of 2 at 10 GeV. The lack of a strong dependence on the halo profile results from the fact that high energy electrons lose energy relatively quickly, leading the observed spectrum to be dominated by particles originating from the surrounding few kiloparsecs.

In calculating the spectrum of electrons and positrons injected through dark matter annihilations, we consider in our analysis three particle physics scenarios:
\begin{itemize}
\item{A 200 GeV WIMP which annihilates to $W^+ W^-$ (which then generate electrons and positrons in their decays). A wino-like neutralino, for example, could annihilate with a large cross section through this channel, although a non-thermal mechanism would be required to populate them in the early universe~\cite{winos}.}
\item{A 600 GeV WIMP which annihilates simply to $e^+ e^-$. Although this case serves primarily as a phenomenological benchmark, models have been constructed in which dark matter annihilates to light states which decay uniquely to electrons and positrons~\cite{arkani} (for a scenario which produces muons in a similar way, see Ref.~\cite{Nomura:2008ru}).}
\item{A 600 GeV Kaluza-Klein dark matter particle in a model with a single flat universal extra dimension~\cite{kkdm}. In this case, the dark matter annihilates to $e^+ e^-$, $\mu^+ \mu^-$, and $\tau^+ \tau^-$ 20\% of the time each, leading to a very hard spectrum of electrons and positrons~\cite{Hooper:2004xn}. Most the remaining annihilations produce up-type quarks.}
\end{itemize}

These three scenarios do not, and are not intended to, cover the full range of phenomenological possibilities for annihilating dark matter. Fits of the PAMELA and ATIC data to a broad range of dark matter masses and annihilation modes have been presented elsewhere~\cite{modelindep}.  Here, we focus our study on how the uncertainties in the propagation model can impact the electron and positron spectrum for the three representative cases described above. In each case, we have used the program PYTHIA~\cite{PYTHIA}, as implemented within DarkSUSY~\cite{DarkSUSY} to calculate the injected spectra of electrons and positrons.

Throughout this study, we normalize the dark matter annihilation rate to a default annihilation cross section of $\langle\sigma v\rangle = 3\times10^{-26}$ cm$^3$/s, multiplied by a ``boost factor''. Such a boost factor could originate, for example, as a result of small scale inhomogeneities in the spatial distribution of dark matter. Alternatively, one could consider non-thermal relics with considerably larger cross sections than our default value, or a thermal relic which annihilates with a very large cross section at low velocities as a result of non-perturbative processes~\cite{sommerfeld}. Variations in the annihilation cross section, boost factor, or average dark matter density lead to changes in the overall normalization of the electron/positron spectrum, but not in the spectral shape.

\begin{figure}
\begin{center}
\epsfig{file=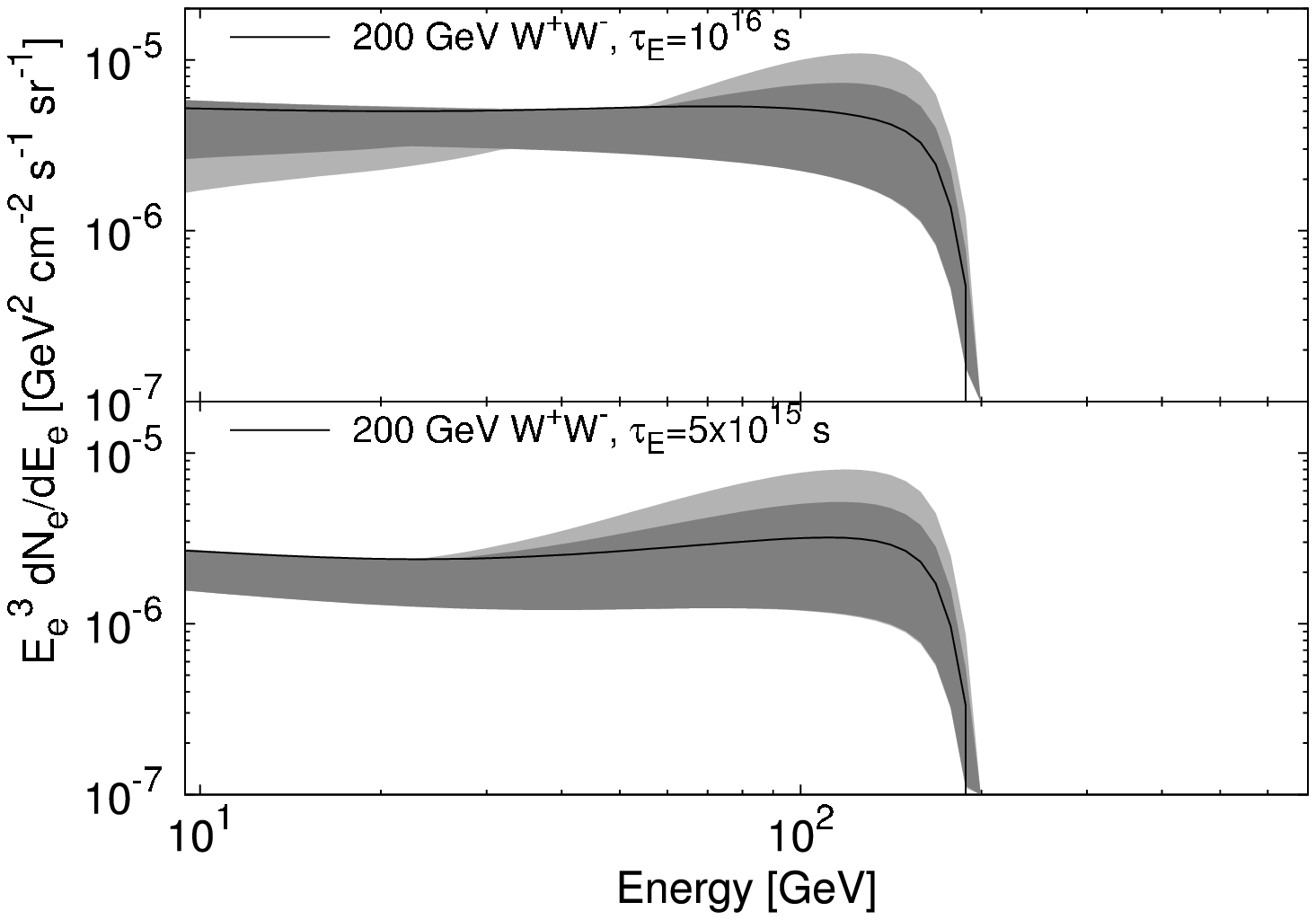,width=0.74\columnwidth}
\epsfig{file=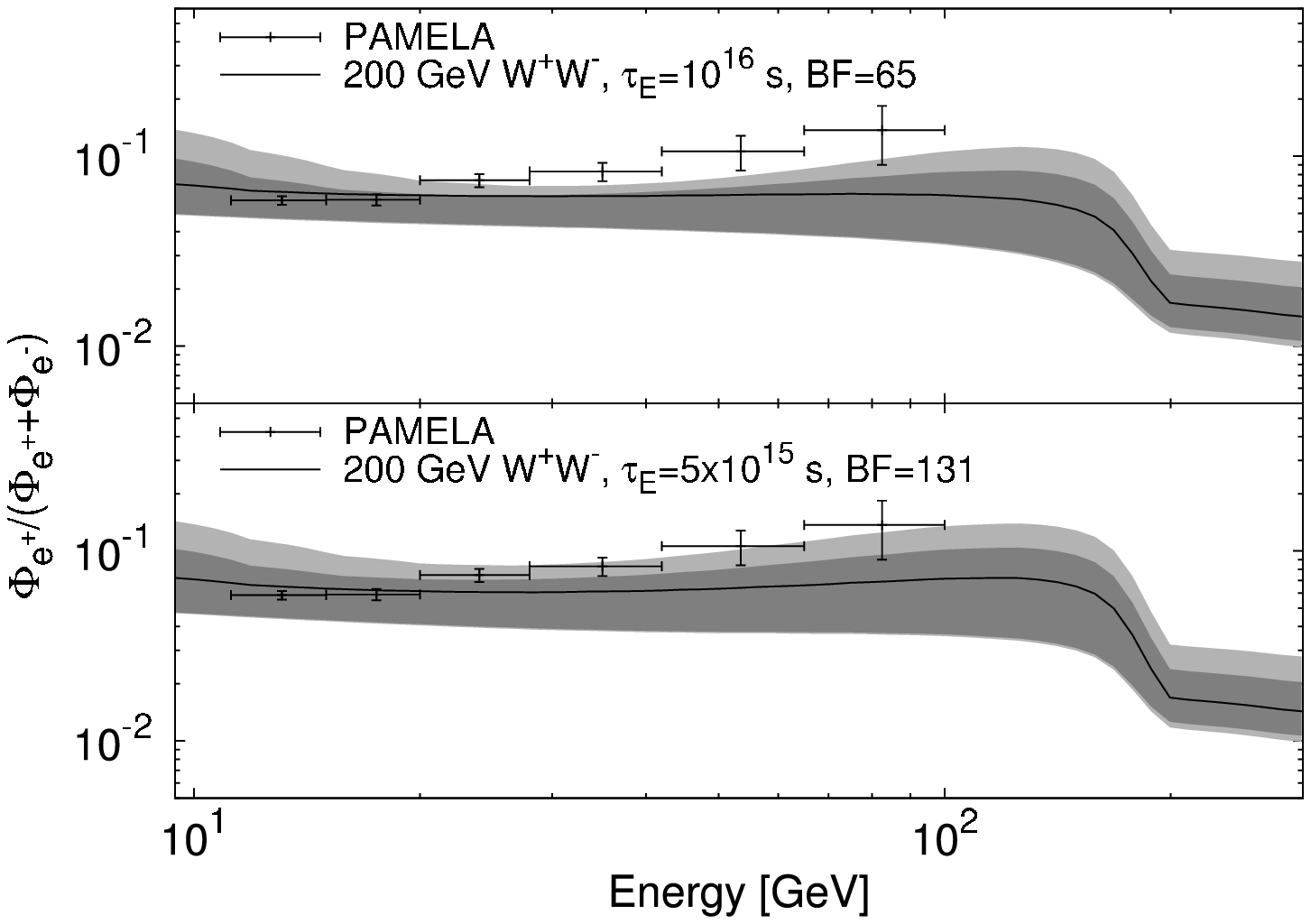,width=0.74\columnwidth}
\end{center}
\vspace{-1.0cm}
\caption{The electron plus positron spectrum (top) and the positron fraction (bottom) predicted from a 200 GeV dark matter particle annihilating to $W^+ W^-$. We show in each upper (lower) frame our results using an energy loss time of $\tau = 10^{16}$ seconds ($\tau = 5 \times 10^{15}$ seconds). For the positron fraction, we compare our results to the PAMELA measurements above 10 GeV. In each frame, the solid line denotes the result using the central propagation model, whereas the dark gray and light gray regions describe the results found over the 1 and 2$\sigma$ range of propagation models (see Fig.~\ref{v0}). We have normalized the annihilation rate to these data using the best fit propagation model (which required boost factors of 65 and 131 for the choice of $\langle\sigma v\rangle = 3\times10^{-26}\,{\rm cm}^3/{\rm s}$). We do not find any set of propagation models consistent with the cosmic ray data that yields a good fit to the PAMELA data for this dark matter model.}
\label{fracWW}
\end{figure}

In Fig.~\ref{fracWW}, we plot the spectrum of electrons and positrons, and the positron fraction, for the case of a 200 GeV particle annihilating to $W^+ W^-$. In the upper and lower frames, we have used two different values for the representative energy loss time, $\tau=10^{16}$ and $5 \times 10^{15}$ seconds, corresponding to radiation and magnetic field densities of $\rho_{\rm{rad}}+\rho_{\rm{mag}} \approx  1$ eV$/$cm$^3$ and $\rho_{\rm{rad}}+\rho_{\rm{mag}} \approx  2$ eV$/$cm$^3$, respectively. As found in previous studies~\cite{modelindep,winospamela}, the positron fraction predicted climbs less rapidly with energy (if at all) than is observed by PAMELA. The set of propagation parameters which provides the best fit to the cosmic ray data discussed in the previous section (shown as solid lines in each frame), yields a very poor fit to the PAMELA data for this dark matter model ($\chi^2/$degree-of-freedom =3.63 and 3.71 for $\tau=10^{16}$ and $5\times 10^{15}$ seconds, respectively). While acceptable variations of the propagation parameters can improve these fits marginally, nothing approaching a good fit can be found.

Even in the unrealistic case that the energy loss rate is reduced to the absolute minimum possible value, corresponding to inverse Compton scattering with the cosmic microwave background alone ($\tau=3.8 \times 10^{16}$ seconds, resulting from a CMB energy density of 0.26 eV/cm$^3$), the positron spectrum from a 200 GeV dark matter particle annihilating to $W^+ W^-$ is too low at high energies (or, if a larger boost factor is used, too high at low energies) to provide a good fit to the PAMELA data~\cite{winospamela}.  Also note that even if this sort of dark matter candidate could potentially generate a signal not very different from that seen by PAMELA, it will not lead to the distinctive feature seen by ATIC.

\begin{figure}[htp]
\begin{center}
\epsfig{file=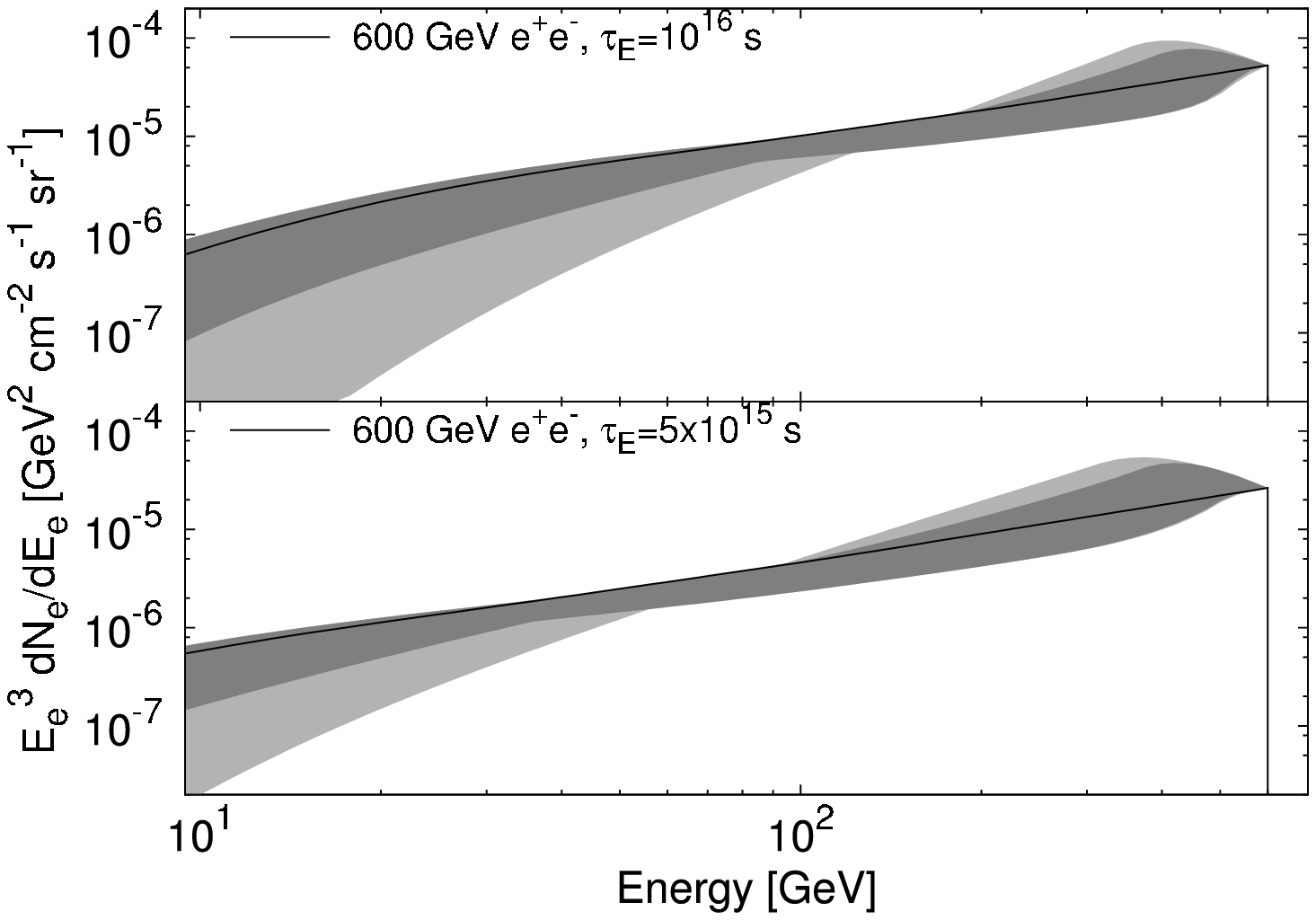,width=0.74\columnwidth}\\
\vspace{0.3cm}
\epsfig{file=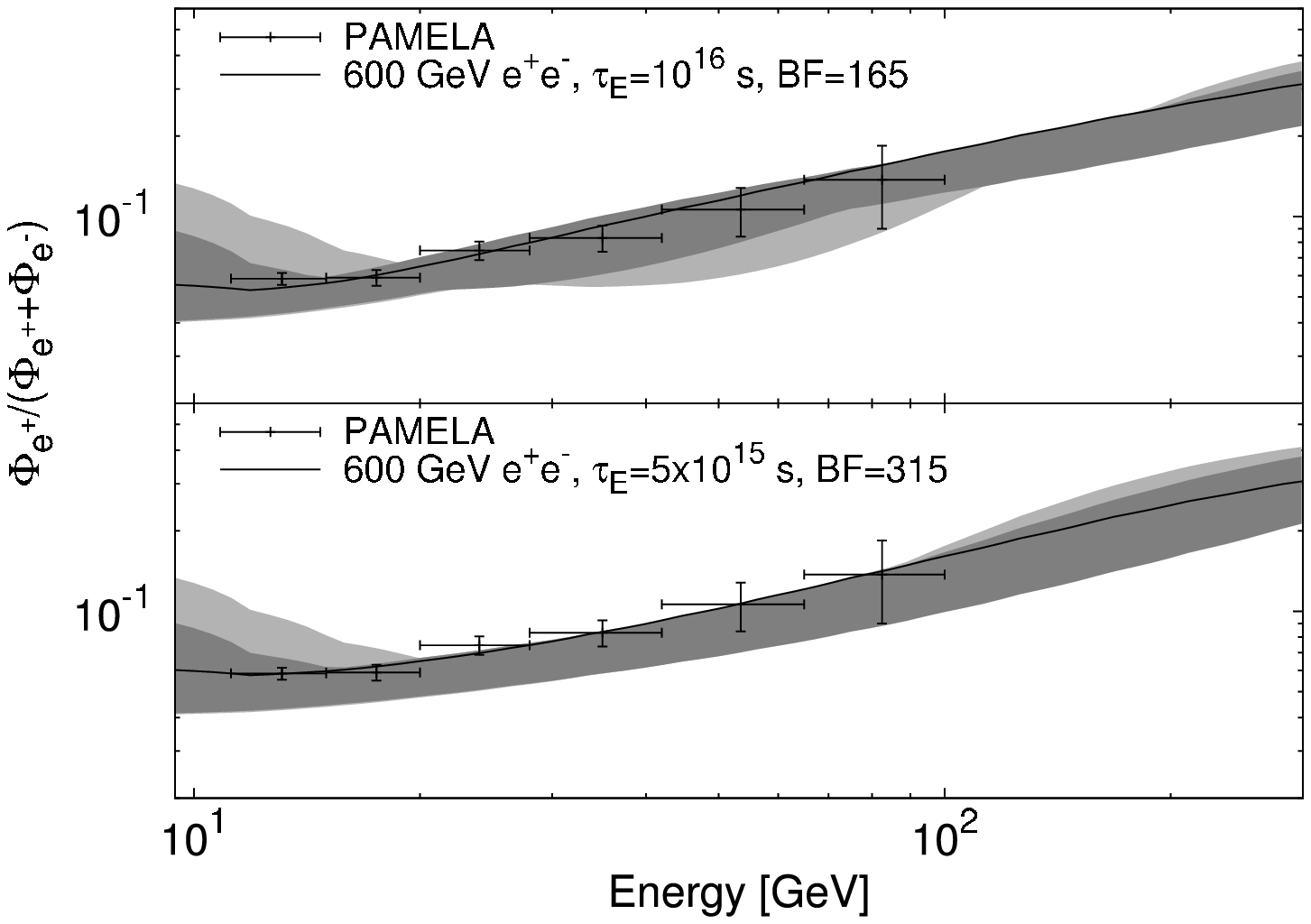,width=0.74\columnwidth}
\end{center}
\vspace{-1.0cm}
\caption{The same as in Fig.~\ref{fracWW}, but for the case of a 600 GeV dark matter particle annihilating to $e^+ e^-$. In this case, boost factors of 165 and 315 were used.}
\label{fracee}
\end{figure}

The inability of a dark matter candidate which annihilates to $W^+ W^-$ to accommodate the positron fraction measured by PAMELA provides us with a motivation to consider dark matter candidates which annihilate directly to electron-positron pairs or to other charged leptons (for examples of scenarios which predict dark matter annihilations to charged leptons, see Ref.~\cite{leptons}). In Fig.~\ref{fracee}, we show the result for the extreme case of a dark matter particle which annihilates uniquely to electron-positron pairs. For this dark matter model, we find that the rapid rise of the positron fraction measured by PAMELA can easily be accommodated. Excellent fits to the PAMELA positron fraction are provided by such a dark matter particle, yielding $\chi^2$ per degree-of-freedom as low as 0.28 and 0.25 for $\tau=10^{16}$ and $5\times 10^{15}$ seconds, respectively. The normalization does require somewhat large boost factors, however, typically within the range of 150 to 350 (for the choice of a 600 GeV mass, selected to accommodate the observed ATIC feature).


\begin{figure}[htp]
\begin{center}
\epsfig{file=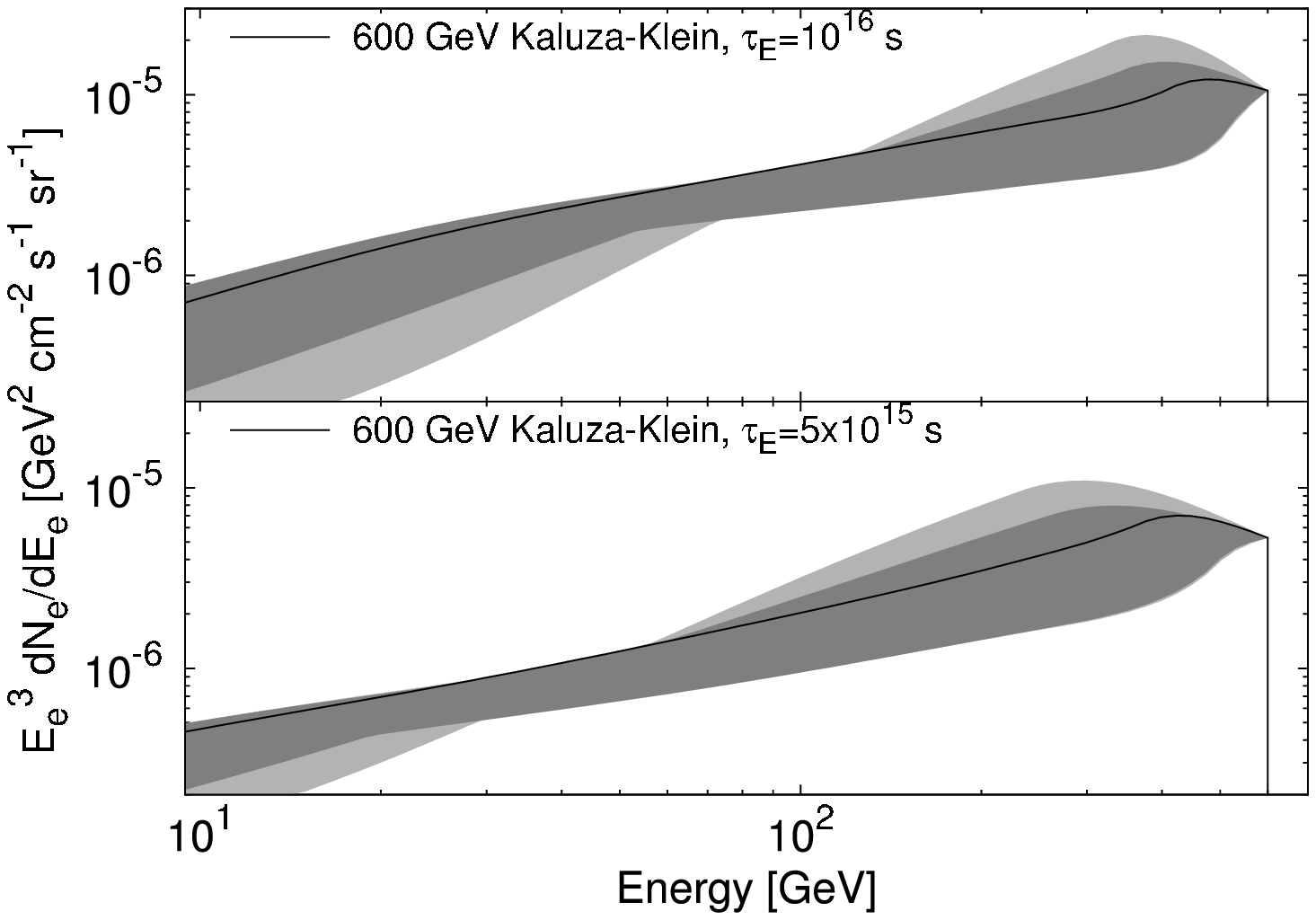,width=0.71\columnwidth}\\
\vspace{0.3cm}
\epsfig{file=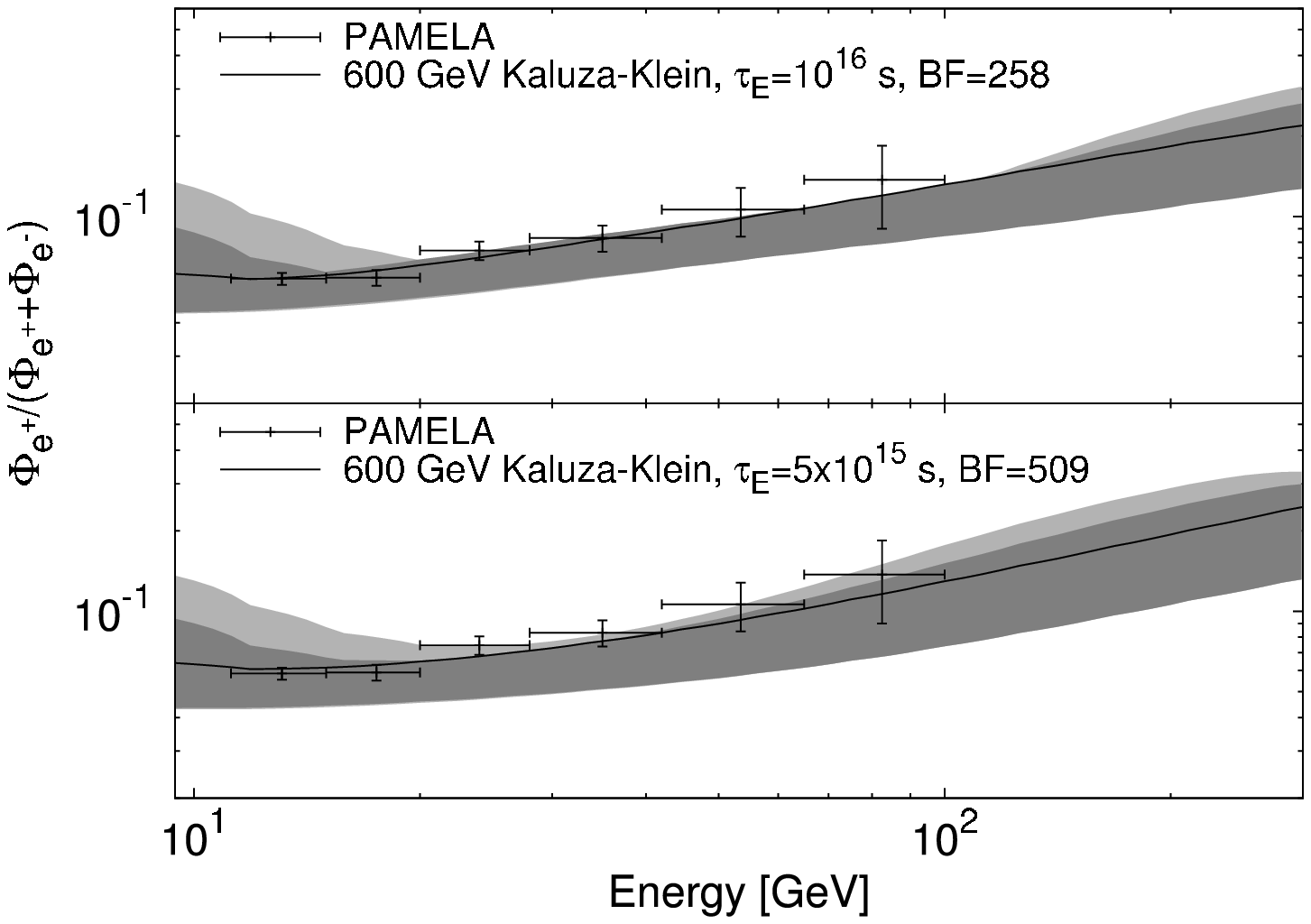,width=0.71\columnwidth}
\end{center}
\vspace{-1.0cm}
\caption{The same as in Figs.~\ref{fracWW} and~\ref{fracee}, but for the case of a 600 GeV dark matter particle annihilating to final states predicted for Kaluza Klein dark matter. Boost factors of 258 and 509 were used.}
\label{fracKKDM}
\end{figure}


In Fig.~\ref{fracKKDM}, we show the results for the case of a 600 GeV Kaluza-Klein dark matter particle, which annihilations to $e^+ e^-$, $\mu^+ \mu^-$, and $\tau^+ \tau^-$ 20\% of the time each. Although the positron fraction grows somewhat less rapidly than in the previous case, the result is still in good agreeement with the measurement of PAMELA. In particular, we find $\chi^2$ per degree-of-freedom as low as 0.34 and 0.51 for $\tau=10^{16}$ and $5\times 10^{15}$ seconds, respectively. Once again, we find that large boost factors are required. Note that in this case, in order to more easily facilitate direct comparison, we continue to use $\langle\sigma v\rangle = 3\times10^{-26}$ cm$^3$/s, although Kaluza-Klein dark matter is actually predicted to have a somewhat higher annihilation cross section.

For the cases of dark matter annihilating to $e^+ e^-$, we expect an edge-like feature to result in the electron-positron spectrum at $E_e=m_X$. The precise spectrum of this feature, however, depends on the detailed distribution of dark matter in the local neighborhood of the Milky Way. This is especially true in the case that $m_X$ is a few hundred GeV or larger. At 600 GeV, for example, an electron will lose most of its energy before traveling much farther than $\sim$1 kpc, leading the high energy spectrum to depend critically on the local dark matter distribution, as well as on the local magnetic structure and other features of the propagation model~\cite{pohl}. For this reason, it is not possible to reliably predict the electron spectrum near $E_e \sim m_X$ with much precision.

\section{Summary and Conclusions}
\label{conclusion}

In this article, we have discussed the propagation of high energy cosmic ray electrons and positrons from dark matter annihilations taking place in the Galactic Halo. We have focused on the astrophysics of this problem, including the properties of the diffusion-energy loss model. To study this process, we have incorporated constraints from measurements of the abundances of stable and unstable secondary nuclei in the cosmic ray spectrum. Stable secondary-to-primary ratios, such as B/C and sub-Fe/Fe, can be used to provide a measurement of the average amount of matter traversed by cosmic rays as a function of energy. Unstable secondary-to-primary ratios, such as $^{10}$Be/$^{9}$Be, can be used to infer the length of time that cosmic rays have been propagating. Taken together, such information can be used to strongly constrain a model of cosmic ray propagation, which can then be used to make predictions for the spectrum of cosmic ray electrons and positrons from a given model of particle dark matter.

We have found that a relatively simple single-zone (cylindrical) diffusion model with a power-law diffusion coefficient and free-escape boundary conditions (and including the effects of diffusive reacceleration, electron K-capture, spallation, and radioactive decay) can describe all of the current cosmic ray data above $\sim$1 GeV. As the precision and quality of this data has improved (in particular, with the latest data from CREAM~\cite{CREAM}), this simple model has held up quite well. At some level, this provides us with confidence that this model constitutes a reasonable description of the processes contributing to cosmic ray propagation in the Milky Way. Although inhomogeneities within the diffusion zone and other considerations will certainly lead to behavior departing from the predictions of this model, the current data do not require any such departures.

Using the propagation parameter sets which we have found to be consistent with the cosmic ray data, we studied the propagation of high energy ($\gsim$$10$ GeV) electrons and positrons in the Galactic Halo. Although considerable variability in these results were found over the acceptable range of propagation parameters, these variations were not sufficient to dramatically change the conclusions reached regarding specific particle dark matter models and the PAMELA and ATIC data. In particular, dark matter annihilating to $W^+ W^-$ throughout the Milky Way halo does not provide a good fit to PAMELA or ATIC data for any acceptable propagation parameter set. In contrast, dark matter which annihilates to electron-positron pairs, or to other charged leptons, naturally provides the rapid climb in the positron fraction as observed by PAMELA (and can also lead to the feature in the electron spectrum observed by ATIC if the mass is chosen to be $\sim$600-800 GeV). We thus conclude that, while variations in the cosmic ray propagation model can lead to substantial variations in the high energy cosmic ray electron and positron spectra from dark matter, we do not expect such variations to lead to qualitative differences to our conclusions regarding the nature of the particle dark matter model required to fit the results of PAMELA or ATIC.

\bigskip

This work has been supported by the US Department of Energy, including grant DE-FG02-95ER40896, and by NASA grant NAG5-10842.

\end{document}